\documentclass[aps,prb,twocolumn,showpacs,groupedaddress]{revtex4}
\usepackage{amsbsy,amssymb,amsmath,bm}
\usepackage{graphicx}

\input{epsf}

\begin{document}

\title {Quantum synchronization in disordered superconducting metamaterials}
\author{M. V. Fistul$^{1,2,3,*}$}
\affiliation {$^{1}$ Theoretische Physik III, Ruhr-Universit\"at Bochum, Universit\"at str. 150, D-44801 Bochum, Germany \\
$^{2}$ National University of Science and Technology MISiS, Leninsky prosp. 4, Moscow, 119049, Russia \\
$^{3}$ Russian Quantum Center, Skolkovo, Moscow region, 143025, Russia\\}

\date{\today}
%\wideabs{
%\maketitle
\begin{abstract}
I report a theoretical study of collective coherent quantum-mechanical oscillations in disordered superconducting quantum metamaterials (SQMs), i.e  artificially fabricated arrays of interacting qubits (two-levels system).  An  unavoidable disorder in qubits parameters results in a substantial  spread of qubits frequencies, and in the absence of electromagnetic interaction between qubits these individual quantum-mechanical oscillations  of single qubits manifest themselves by a large number of small resonant drops in the frequency dependent transmission of electromagnetic waves propagating through disordered SQM, $D(\omega)$. We show that even a weak electromagnetic interaction between adjacent qubits can overcome the disorder and establish completely or partially  \emph{synchronized} quantum-mechanical dynamic state in the disordered SQM. In such a state a large amount of qubits displays the collective quantum mechanical oscillations, and this collective behavior manifests itself by a few giant resonant drops in the $D(\omega)$ dependence. The size of a system $r_0$ showing the collective (synchronized) quantum-mechanical behavior is determined in the one-dimensional SQMs as $r_0 \simeq a [K/\delta \Delta]^2$, where $K$, $\delta \Delta$, $a$ are the energy of nearest-neighbor interaction, the spread of qubits energy splitting, and the distance between qubits, accordingly. We show that this phenomenon has an origin in the Anderson localization of spinon-type excitations arising in the SQM. Our analysis is also in a good accord with recent experiments on the electrodynamics of the disordered 1D SQMs.
%Ref. \cite{SQM6}.
\end{abstract}


\maketitle

\section*{Introduction}
%\emph{Introduction }
Superconducting quantum metamaterials (SQMs) are novel artificially prepared solid state structures whose electrodynamic properties are governed by  peculiar interplay of  classical Maxwell electrodynamics and  quantum-mechanical laws \cite{SQM6,SQM1,RZ,SQM2,SQM3,SQM4,SQM5,KMS,Tsironis}. Most of SQMs  consists of an array of interacting superconducting qubits (two-level systems) , e.g. charge qubits \cite{ChQubit}, flux qubits \cite{Flqubit}, transmons \cite{TRqubit} etc.

Numerous experiments on the propagation of electromagnetic waves through the SQM \cite{SQM6,SQM3,SQM4,SQM5} have allowed one to observe an excitation of  coherent quantum-mechanical oscillations (beatings) between two states of single qubits.  This intrinsically quantum-mechanical macroscopic phenomenon manifests themselves by sharp resonant drops in the frequency dependent transmission coefficient $D(\omega)$ of electromagnetic waves  propagating in the transmission line electromagnetically (inductively or capacitively) coupled to the SQM \cite{SQM6,SQM3,SQM4,SQM5,TRqubit,FistVolkov,UstTheoryDisorder,UstFistSchulga}. Such measurement setup is presented schematically in Fig. 1.
%, embedded in a low-dissipative resonator.
%A large amount of diverse coherent macroscopic quantum-mechanical phenomena such as coherent quantum oscillations between two macroscopic states of %single qubits, microwave induced Rabi oscillations, Ramsey fringes, just to name a few, have been observed in SQMs. Moreover,  the macroscopic %quantum dynamics of SQMs can provide both the Kerr type of non-linearity  and non-linear microwave pulse (soliton)  propagation  in the SQMs.
In a simplest case of an "ideal" SQM consisting of an array of \emph{non-interacting identical} qubits the $D(\omega)$ displays a single drop on the frequency $\omega=\Delta/\hbar$, where each qubit is characterized by the energy level differences $\Delta$. A small width of such drops is determined by various dissipative and decoherent processes.

However, in imperfect SQMs an  unavoidable disorder in superconducting qubits parameters is always present, and it results in a substantial spread of  energy level differences $\Delta_i$. In this case it is plausible to suppose that the \emph{individual} coherent quantum-mechanical oscillations of single qubits occur in the SQM, and the $D(\omega)$ has to display a large amount of small resonant drops up to the value $N$, where $N$ is a number of qubits in the SQM \cite{FistVolkov,UstTheoryDisorder}. Indeed, in Ref. \cite{SQM5} the frequency dependent transmission coefficient $D(\omega)$ of the SQM containing $7$  non-interacting qubits has been measured, and $7$ different small resonant drops in the $D(\omega)$ have been observed. Surprisingly in other experiments carried out on diverse SQMs a small number (one or two) of giant  resonant drops has been observed in spite of the presence of  large number of qubits with different parameters \cite{SQM6,UstFistSchulga}. However, the physical conditions necessitating   to observe such \emph{collective (synchronized) quantum-mechanical behavior }are not clear at this moment.

A key point allowing one to resolve such a puzzle is \emph{electromagnetic interactions} between qubits. Two types of interactions can be realized in the SQMs : the nearest-neighbor interaction that is due to direct inductive or capacitive electromagnetic interaction between adjacent qubits, and/or the long-range electromagnetic interaction arising due to consequent emission, propagation and absorbtion of virtual photons in low-dissipative resonator \cite{FistVolkov,Fist}.

In this Article taking into account the inductive interaction between adjacent qubits we show that the interaction between qubits can overcome the disorder and establish collective synchronized quantum-mechanical oscillations characterized by just few frequencies. We obtain that even in the case of a weak interaction, i.e.  $K \ll \Delta_i$ ($K$ is the typical energy of nearest-neighbor interactions), a large amount of qubits $N^\star \simeq (K/\delta \Delta)^2$ are synchronized, and these qubits display collective quantum-mechanical oscillations of a single frequency. Correspondingly, $N/N^\star$ is a total number of diverse quantum oscillations observed in disordered SQMs. Here, $\delta \Delta$ is the typical spread of energy level differences of individual qubits.  Such synchronization phenomenon has an origin in the Anderson localization \cite{AL} of spinon-type excitations occurring in the SQM.

\section*{Results}

\subsection*{Superconducting Quantum Metamaterial Model and Hamiltonian of disordered SQM.}

Let us to consider a one-dimensional array of $N$ qubits modeled as two-level systems, with nearest-neighbor inductive electromagnetic interactions between adjacent qubits.  In a particular case of flux qubits ($3$-Josephson junction SQUID) the two states correspond to the clockwise and anticlockwise flowing currents, and the energy level difference, $\Delta_i$, is determined by quantum tunneling between these states \cite{Flqubit}. The frequencies of individual coherent quantum-mechanical oscillations that can be excited in a single qubit, are $\omega_i=\Delta_i/\hbar$. The SQM is coupled to the linear transmission line, and this setup allows one to study the propagation and reflection of electromagnetic waves through the SQM.  The schematic of such setup is shown in Fig. 1. Notice here, that similar setup has been used in Ref. \cite{SQM6} in order to measure macroscopic quantum-mechanical oscillations excited in the disordered SQM.

The quantum dynamics of each qubit is characterized by the imaginary-time dependent degree of freedom, $\varphi_i (\tau)$, and the qubits Hamiltonian has a form:
  \begin{equation} \label{Hamiltonian of qubits}
H_{qubits}=\sum_{i=1}^{N} \frac{m^\star}{2}[\dot{\varphi}_i]^2+U_i[\varphi_i],
\end{equation}
where both the parameter $m^\star$ and  the potentials $U_i[\varphi]$ determine completely quantum-mechanical dynamics and  the energy levels of isolated qubits. Moreover, the double-well potential $U_i[\varphi]$  results in  small quantum-mechanical energy level differences $\Delta_i$ of single qubits. The unavoidable disorder in qubit parameters leads to a spread of $\Delta_i$ in the SQM. The Hamiltonian of inductive electromagnetic interaction between adjacent qubits is written as
  \begin{equation} \label{InterHamiltonian}
H_{int}=\sum_{i=1}^{N} \frac{K}{2}[\varphi_i-\varphi_{i-1}]^2,
\end{equation}
where the strength of nearest-neighbor interaction, $K$, is determined by mutual inductance between qubits.

\begin{figure}[tbp]
\includegraphics[width=0.9in,angle=-90]{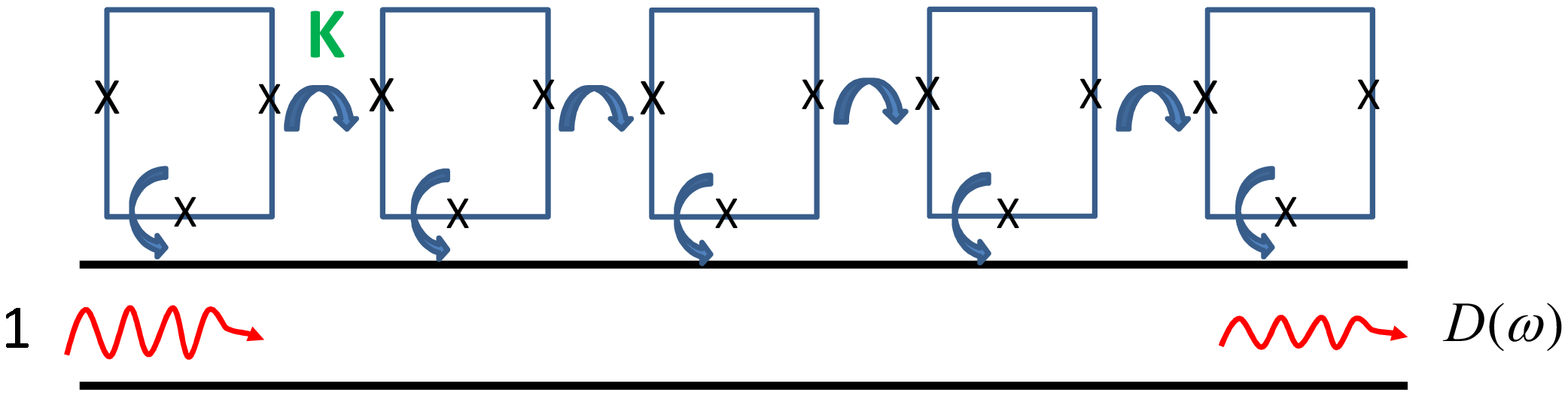}
\caption{\textbf{The schematic of an experimental setup allowing one to observe the coherent quantum-mechanical oscillations in the SQM}.  The SQM based on an array of flux qubits ($3$-Josephson junction SQUID) is shown. The array of qubits is coupled to the transmission line . The inductive coupling between adjacent qubits, $K$, is taken into account.
} \label{schematic}
\end{figure}

\subsection*{Partition Function and Instanton Analysis} The thermodynamic properties of disordered SQMs are determined by the partition function $Z$ expressed through the Feynman path integral in the imaginary-time $\tau$ representation as
  \begin{equation} \label{partition function}
Z=\int D[\varphi_i] \exp \left[-\frac {1}{\hbar}\int_0^{\frac{\hbar}{k_B T}}d\tau H \{\varphi_i \} \right], ~H=H_{qubit}+H_{int}.
\end{equation}

We consider disordered SQMs with  a weak electromagnetic interaction, i.e. $K \ll \Delta_i$, and in this case the optimal configurations in the path integral are  series of alternating instanton (anti-instanton ) solutions uncorrelated in imaginary time interval $[0, \hbar/(k_B T)]$ \cite{FistVolkov,KivChack} ( schematic of a typical path configuration is shown in Fig. 2).  As the electromagnetic interaction between qubits is absent, i.e. $K=0$, one can obtain $Z=\prod_{i=1}^N \cosh [\Delta_i/(2k_B T)]$, and on the time interval $[0, \hbar/(k_B T)]$ the average quantity of instantons and anti-instantons for an $i$-qubit $<N_i>=\Delta_i/(k_B T)$ \cite{KivChack}. To analyze the collective behavior of disordered SQM we will characterize each qubit by  random value of $N_i$, and the probability $P_{K=0}\{N_i \}$ shows  sharp peaks on the values $N_i=<N_i>$ as
\begin{equation} \label{ProbabilityK=0}
P_{K=0}\{N_i \} \propto \prod_{i=1}^N \exp \left \{ \frac{k_BT }{2\Delta_i}[N_i-\Delta_i/(k_B T)]^2 \right \}.
\end{equation}

\begin{figure}[tbp]
\includegraphics[width=1.7in,angle=-90]{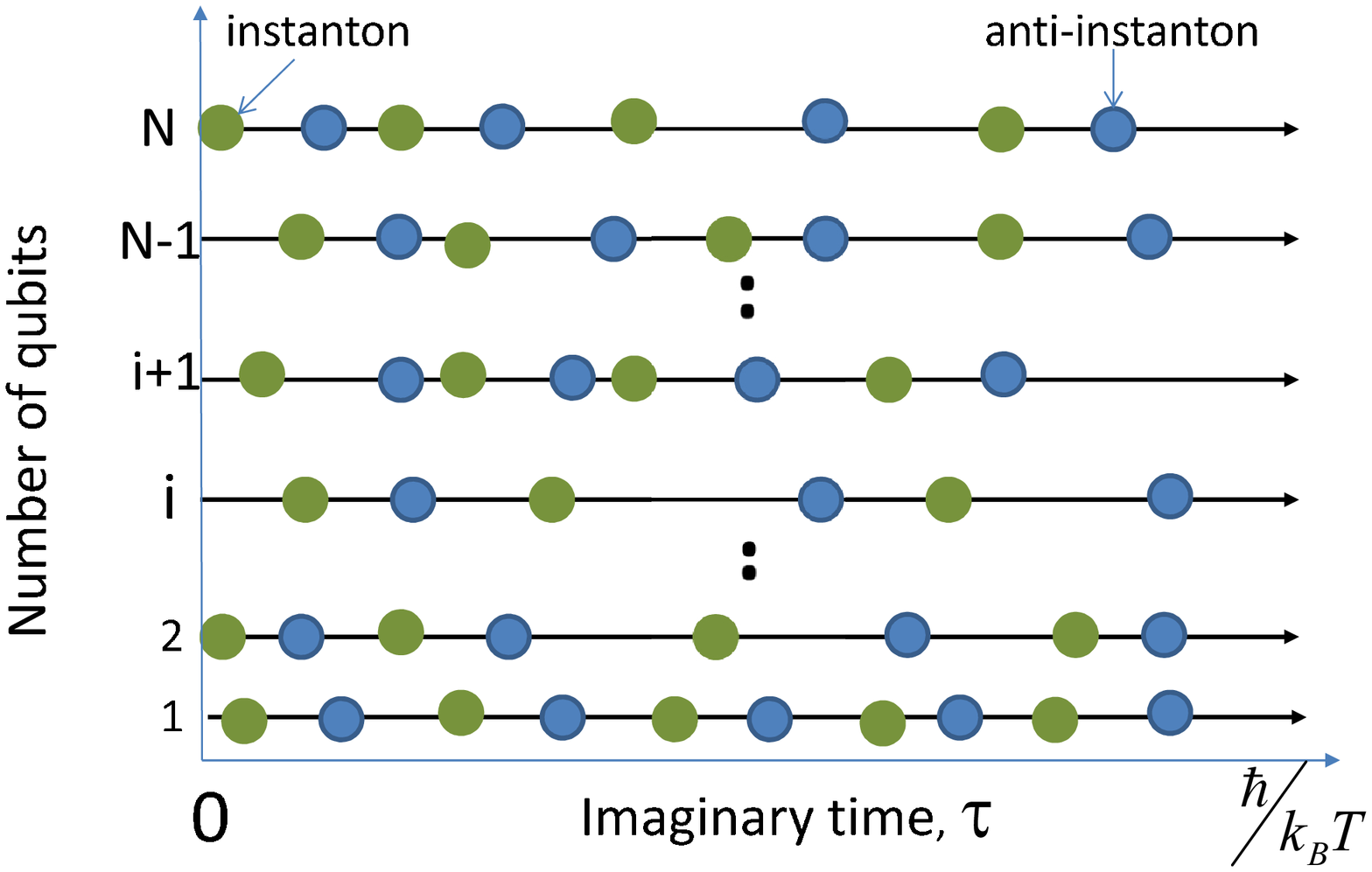}
\caption{\textbf{The schematic of a typical path configuration} It consists of alternating instanton (green circles) and anti-instanton (blue circles) solutions. The $\tau$ is the imaginary time. Each qubit is characterized by a total quantity of instantons and anti-instantons, $N_i$.
} \label{schematic-2}
\end{figure}

Next, we turn to the analysis of coherent quantum mechanical oscillations in disordered SQMs with a weak ($K \ll \Delta_i$) intrinsic electromagnetic interaction between qubits . Substituting instanton-(anti)instanton solutions in the Eqs. (\ref{partition function}) and (\ref{InterHamiltonian}) and introducing the qubits frequencies as $\omega_i=k_BT N_i/\hbar $ we obtain the probability $P_{K} \{\omega_i \}$ in the following form:
\begin{eqnarray}
   P_{K}\{\omega_i \} =\frac{1}{Q} \int D \{\omega_i \} \prod_{i=1}^N \exp \left \{ -\frac{\hbar \tilde{K}}{k_B T \Delta}|\omega_i-\omega_{i-1}| \right \} \times
    \nonumber \\
    \times \exp \left \{-\frac{\hbar^2}{2k_B T \Delta}[\omega_i-\Delta_i/\hbar]^2 \right \}\label{ProbabilityK}
\end{eqnarray}
and the normalizing constant $Q$ is expressed through the path integral as
\begin{eqnarray}
   Q =\int D\{\omega_i \}\prod_{i=1}^N~~~~~~~~~~~~~~~~~~~~~~~~~~~~~~~~~~~~~~~
    \nonumber \\
  \exp \left \{ -\frac{\hbar \tilde{K}}{k_B T \Delta}|\omega_i-\omega_{i-1}|-\frac{\hbar^2 }{2 k_B T \Delta}[\omega_i-\Delta_i/\hbar)]^2 \right \}. \label{Normvalue}
\end{eqnarray}
Here, $\Delta=\bar{\Delta}_i$ is the average value of energy splitting of qubits $\Delta_i$, and $\tilde{K}=K \varphi_0^2/2$ is the effective strength of electromagnetic interaction between instantons (anti-instantons) of adjacent qubits; $\varphi_0$ is the the value of $\varphi$ in the classical stable state.  We have taken into account that the spread of qubit splitting is  small, i.e. $\delta \Delta \ll \Delta $.
The physical meaning of different terms in Eq. (\ref{Normvalue}) is rather transparent: the disorder results in a spread of instanton quantities, $N_i$ along the array, but the electromagnetic interaction  allows to \emph{equalize} the quantity of instantons on different qubits. Since  the frequencies of quantum-mechanical oscillations  are directly related to the quantities of instantons, one expect that the electromagnetic interaction results in the synchronization phenomena.
%Notice here, that the interaction between instanton liquids of different qubits differs from an usual diffusive term

\subsection*{Correlations of Qubits Frequencies}

To analyze this effect quantitatively we consider the continuous limit as $N \gg 1$, and using the periodic boundary conditions we expand $\omega (x)$ and $\Delta (x)$ ($x$ is the coordinate along the array of qubits) in Fourier series as:
\begin{eqnarray}
\omega (x)=\frac{\Delta}{\hbar}+\frac{1}{\sqrt{L}}\sum_{n=-\infty}^{\infty}  ~a_n \exp \left [ i\frac{2\pi n x}{L} \right ],
    \nonumber \\
\Delta (x)= \Delta+\frac{1}{\sqrt{L}}\sum_{n=-\infty}^{\infty}  ~b_n \exp \left [ i\frac{2\pi n x}{L} \right ] \label{Fouriertransform}
\end{eqnarray}
Here, $L=Na$ is the length of the SQM, and $a$ is the distance between qubits.

Substituting (\ref{Fouriertransform}) in (\ref{Normvalue}) we obtain the partition function of interacting instanton liquid, $Q$,   in the following form:
\begin{eqnarray}
 Q =\int D\{a_n \}\exp \left \{ -\frac{\hbar \tilde{K}}{k_B T \Delta}\sqrt{L \sum_n |a_n|^2 (\frac{2\pi n}{L})^2 } \right \}
    \nonumber \\
\exp \left \{ -\frac{\hbar^2 }{2k_B T \Delta a}\sum_n |a_n-b_n/\hbar|^2 \right \}.  \label{Normvalue-1}
\end{eqnarray}
Transforming the Eq. (\ref{Normvalue-1}) to the following form
\begin{eqnarray}
Q =\int_0^{\infty} dX \int_{-\infty}^{\infty} dY \int D\{a_n \} ~~~~~~~~~~~~~~~~~~~~~~~~~~~~~~~~~~~~~~\nonumber \\
\exp \left \{ -\frac{\hbar \tilde{K}}{k_B T \Delta }\sqrt{ L X} +iXY-iY \sum_n |a_n|^2 (\frac{2\pi n}{L})^2 \right \} \times \nonumber \\
\times \exp \left \{ -\frac{\hbar^2 }{2 k_B T \Delta a} \sum_n |a_n-b_n/\hbar|^2 \right \}. ~~~~~~~~~~~~~~~~~~~~~~~~~~ \label{Normvalue-2}
\end{eqnarray}
and calculating the Gaussian integrals over $a_n$ we obtain
\begin{eqnarray}
Q = \int_0^{\infty} dX \int_{-\infty}^{\infty} dY \exp \left \{ -\frac{\hbar \tilde{K}}{k_B T \Delta }\sqrt{ L X} +iXY \right \}
    \nonumber \\
\exp \left \{\sum_n \frac{|b_n|^2 }{ 2k_B T \Delta a [\frac{2 k_B T \Delta a }{\hbar^2} iY(\frac{2\pi n}{L})^2+ 1 ] } \right \}. \label{Normvalue-3}
\end{eqnarray}
Optimal values of $X_0$ and $Y_0$ are obtained from systems of equations as:
\begin{eqnarray}
\frac{\hbar \tilde{K}\sqrt{L}}{2 k_B T \Delta \sqrt{ X_0}}=iY_0~~~~~~~~~~~~~~~~~~~~~~~~~~
    \nonumber \\
X_0=\frac{1 }{\hbar^2} \sum_n \frac{|b_n|^2 (\frac{2\pi n}{L})^2 }{[\frac{2 k_B T \Delta a }{\hbar^2} iY_0(\frac{2\pi n}{L})^2+ 1 ]^2 }
\label{SystemEqs}
\end{eqnarray}
Taking into account that the values of $\Delta_i$ are not correlated for different qubits we obtain $|b_n|^2=2(\delta \Delta)^2 a$.
Therefore, one can obtain $iY_0=4\hbar^2 \tilde{K}^4 a/[2k_B T \Delta (\delta \Delta)^2 ]$.

In spite of presence of uncorrelated disorder in qubits splitting $\Delta_i$ the electromagnetic interaction between qubits can lead in long-range correlations of frequencies of quantum-mechanical oscillations excited in the disordered SQMs. These correlations are described quantitatively by coordinate-dependent correlation function of qubit frequencies:
\begin{eqnarray}
<\omega(x_1)\omega (x_2)>=(\Delta/\hbar)^2+R(x_1-x_2) \nonumber \\
R(x_1-x_2)=\frac{1}{L} \sum_n |a_n|^2 \exp \left [ i\frac{2\pi n (x_1-x_2)}{L} \right ].
\label{CorrelFunction}
\end{eqnarray}
The amplitudes $a_n$ are determined by random values of qubit splitting $b_n$ as $a_n=(1/\hbar)  b_n [(2\pi n r_0/L)^2+1]^{-1}$, where $r_0^2 =2k_B T  \Delta a iY_0/\hbar^2 $, and $r_0$ is the correlation radius determining the area where the quantum-mechanical oscillations are synchronized. The correlation function $R(x)$ is written explicitly as
\begin{eqnarray}
R(x)=\frac{a}{2r_0} \frac{(\delta \Delta)^2}{\hbar^2}\left (  1+ \frac{x}{r_0} \right ) \exp [-|x|/r_0],
\label{CorrelFunction-2}
\end{eqnarray}
where the correlation radius $r_0=2a [\tilde{K}/(\delta \Delta)]^2$ can greatly exceed the distance between qubits $a$. A peculiar dependence of correlation radius on the strength of disorder has an origin in a  sub-diffusive interacting term $ \propto |\omega_i-\omega_{i-1}|$ in the exponent of Eq. (\ref{Normvalue}). It differs such a problem from e.g. fluctuation induced bending of strings and superconducting vortex lines \cite{VinNelson}.

Next, we notice that this analysis can be extended to the two-dimensional  lattice of interacting qubits. Indeed, the Eqs. (\ref{SystemEqs}) are valid for $2d$ square lattice with the substitution: $(1/L^2)\sum_n \rightarrow \int q dq $. Calculating all integrals we obtain that in a 2d case the correlation radius is exponentially large, i.e. $r^{(2)}_0 ~\simeq a \exp(2 [\tilde{K}/(\delta \Delta)]^2)$.

The intrinsic correlations of frequencies, and, in particular, the dependencies of the correlation radius $r_0$ on the strength of interaction $\tilde{K}$ and the disorder $\delta \Delta$ strongly resemble the phenomenon of Anderson localization in disordered low-dimensional solid state systems \cite{AL}. The origin of such similarity is the following: the disordered SQMs with nearest-neighbor interactions are mapped exactly on the quantum Izing model with a large transverse magnetic field \cite{spinmodel1,Levitov}. The Hamiltonian of such a  model is written as
\begin{eqnarray}
\hat{H}_{spin}=\sum_{i,j} \Delta_i \hat{s}^{(i)}_{x}+\tilde{K} \hat{s}^{i}_z\hat{s}^{j}_z \label{SpinHamiltonian}.
\end{eqnarray}

In the limit of $\tilde K < \Delta$ the ground state of this model is the spin ordering in the $x$ direction, and the low-lying excited states form the  \emph{spinon band.} separated from the ground state by the energy gap of order $\Delta$. In the presence of disorder all spinon states are localized, and a spread  of localized wave function of spinons is of order $r_0$. Such Anderson localization in the spinon band is shown  schematically in Fig 3. Therefore, the coherent quantum mechanical oscillations correspond to the resonant transitions  between the ground state and the various excited states of localized spinons.

\begin{figure}[tbp]
\includegraphics[width=3in,angle=0]{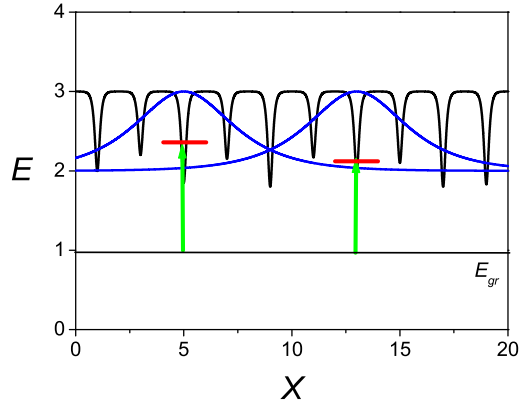}
\caption{\textbf{The schematic of the Anderson localization in the quantum Izing model with transverse magnetic field.} The ground state, the effective potential for the low-lying excited states (the spinon band) are shown. The wave functions of excited states (blue lines) are localized, and a spread of wave functions characterizes These oscillations correspond to the resonant transitions (green lines) between the ground state and the excited states of localized spinons.
} \label{spinmodel}
\end{figure}

\subsection*{Electromagnetic Waves Transmission through the SQM.}

The coherent quantum-mechanical oscillations excited in the SQM can be observed through the frequency dependent transmission coefficient $D(\omega)$ of electromagnetic waves propagating in the transmission line inductively coupled to the SQM \cite{SQM6,UstFistSchulga}. The $D(\omega)$ is determined by time-dependent quantum-mechanical correlation functions of qubits $C(t)=\sum_{i=1}^N C_i(t)$ \cite{FistVolkov,UstTheoryDisorder,Ingold}. The correlation function $C_i(t)$ displays oscillations of frequency $\omega_i$, and taking into account the finite dissipation $\gamma$ one can write down a generic expression for the transmission coefficient $D(\omega)$ as a set of resonant drops
  \begin{equation} \label{Transmission coefficients}
D(\omega)=1-\sum_{i=1}^N \frac{\alpha}{\left (\omega-\omega_i \right)^2+\gamma^2},
\end{equation}
where the parameter $\alpha \ll \gamma^2$  characterizes  the coupling between the transmission line and SQM. Thus, as the electromagnetic interaction between adjacent qubits are small, i.e. $ K \ll \delta \Delta $, and a substantial spread of qubit splitting $\delta \Delta \gg \gamma$ , the disordered SQM supports $N$ non-synchronized coherent quantum oscillations of different frequencies. These non-synchronized coherent quantum-mechanical oscillations manifest themselves by $N$ small resonant drops in the $D(\omega)$ (see, Eq. (\ref{Transmission coefficients})).

However, the crossover to partially synchronized regime occurs as the inductive interaction between nearest-neighbors  qubits overcomes the disorder in qubits energy level differences, i.e. $K \geq \delta \Delta$. In this regime  the correlation radius $r_0$  exceeds the distance $a$ between qubits, and a large amount of qubits, $N^\star \simeq r_0/a \gg 1$ displays  collective synchronized behavior characterized by a single frequency. As the interaction $K$ goes over the value $(\delta \Delta) N^{1/2}$ all qubits become synchronized, and the collective quantum mechanical oscillations are established in a whole SQM. Notice here, that  main assumption of our analysis is the absence of correlations of instanton (anti-instanton) positions on the $\tau$-axis, and this assumption is valid as $K < \bar{\Delta}$.

The fingerprints of synchronized regimes of coherent quantum oscillations  are a few number of giant resonant drops in the frequency dependent electromagnetic waves transmission, $D(\omega)$ (see Fig. 1, and Eq. (\ref{Transmission coefficients})). In the synchronized quantum-mechanical dynamic state  a number of such resonant drops is $N/N^\star=L/r_0= (N/2) [\delta \Delta/K]^2$. In Ref. \cite{SQM6} the propagation of electromagnetic waves through the SQM containing of 20 qubits has been experimentally studied. In these experiments one or two giant resonant drops in the frequency dependent transmission coefficient, $D(\omega)$, have been observed. These results indicate the excitation of synchronized coherent quantum mechanical oscillations in the SQM, and we argue that this novel state occurs due to the presence of substantial inductive coupling between adjacent qubits, i.e. $K /(\delta \Delta) \simeq 2.5-3$.

However, one can not exclude an alternative explanation that a long-range interaction  in an array of qubits also can induce the synchronized quantum-mechanical dynamic state.  Such an interaction originates from  the emission (absorption) of virtual photons of a low-dissipative resonator \cite{SQM1,FistVolkov,Fist}. This type of interaction  leads to the interaction between instantons (anti-instantons) of different qubits, and it also allows to equalize quantities of instantons (anti-instantons) on different qubits, and therefore, to establish a synchronized regime. We will address the quantitative analysis of synchronized quantum-mechanical oscillations induced by long-range interaction, elsewhere.

\section*{Conclusions} In conclusion we have theoretically analyzed the excitation of  coherent quantum-mechanical oscillations in disordered SQMs. Our analysis is based on the mapping of coherent quantum-mechanical oscillations to series of alternating instanton (anti-instanton) solutions in the path-integral approach.
In this model the frequencies of quantum-mechanical oscillations are directly related to the quantity of instantons on different qubits, $N_i$. The disorder results in a spread of $N_i$ along the array of qubits, and a weak electromagnetic interaction between adjacent qubits, $K$, leads to the partial alignment of these quantities, $N_i$. Thus, we have obtained that a large amount of qubits can display synchronized collective behavior characterized by a single frequency. The size of the region showing synchronized behavior is determined by the ratio of  the strength of  interaction $K$ to the  typical spread of energy splitting of qubits, $\delta \Delta$.

Therefore, the fabrication of disordered SQMs with a tunable inductive coupling allows one to  observe the crossover from a non-synchronized regime to the synchronized regime , and provides the direct evidence of synchronized quantum-mechanical oscillations.  We anticipate that the realization of synchronized regime in intrinsically disordered SQMs will result in a substantial simplification of the process of qubits addressing. Finally, since arrays of interacting qubits are exactly mapped on various quantum spin models, our method of instanton analysis can be used for a quantitative theoretical study of low-lying quantum-mechanical states in diverse interacting disordered quantum systems under equilibrium and non-equilibrium conditions \cite{spinmodel1,Levitov,spinmodel2,Nori}.

\textbf{Acknowledgments}
I would like to thank A. V. Ustinov and A. Sedrakyan for useful discussions. This work was supported by the Russian Science Foundation (grant No. 16-12-00095).
I acknowledge 
%from the Ministry of Education and Science of Russian Federation in the frame of Increase Competitiveness Program of the NUST MISIS (contracts no. K2-2014-015) 
the hospitality of the International Institute of Physics, Natal Brazil where this work has been finished.

{}

\end{document}